\documentclass{PoS}

\usepackage{mathtools}
\usepackage{xcolor}
\usepackage[most]{tcolorbox}
\usepackage{graphicx} 
\usepackage{booktabs} 
\usepackage{amsmath,bm}
\usepackage{subfig}

\title{Exploring New Physics in {\bm{$B \rightarrow \pi K$}} Decays}

\ShortTitle{Exploring New Physics in {{$B \rightarrow \pi K$}} Decays}

\author{\speaker{Eleftheria Malami}\\
        Nikhef, Science Park 105, NL-1098 XG Amsterdam, Netherlands\\
        E-mail: \email{emalami@nikhef.nl}}

\abstract{B-meson decays play an important role in flavour physics. The $B \rightarrow \pi K$ decays are dominated by QCD loop diagrams (penguins) but also electroweak penguins, where New Physics may enter, have a significant impact on the decay amplitude. Since measurements from B-factories indicate deviations from the Standard Model picture, we perform a state-of-the-art analysis to explore the correlation of the CP asymmetries and to get an updated picture. We propose a strategy for the optimal determination of the parameters which describe electroweak penguin effects and apply it to current data, utilising both neutral and charged $B \rightarrow \pi K$ decays. This new method can be fully exploited at the future Belle-II experiment, which will hopefully answer the question: Do these decays imply New Physics?}

\FullConference{Corfu Summer Institute 2019 "School and Workshops on Elementary Particle Physics and Gravity" (CORFU2019)\\
		31 August - 25 September 2019\\
		Corfù, Greece}

\begin{document}

\section{Introduction}
The B-meson system is a very important system to test the flavour sector of the Standard Model (SM), to explore CP violation, which within the SM is described by the Cabibbo-Kobayashi-Maskawa (CKM) matrix \cite{Cab,Kob}, and to gain more information as far as physics beyond the SM is concerned. Particularly interesting channels for such studies are the $B \rightarrow \pi K$ decays (for a selection of references, see \cite{NQ}-\cite{London}).

\begin{figure}[b!]
	\centering
	\subfloat[]{\label{fig:1}\includegraphics[width = 0.35\linewidth]{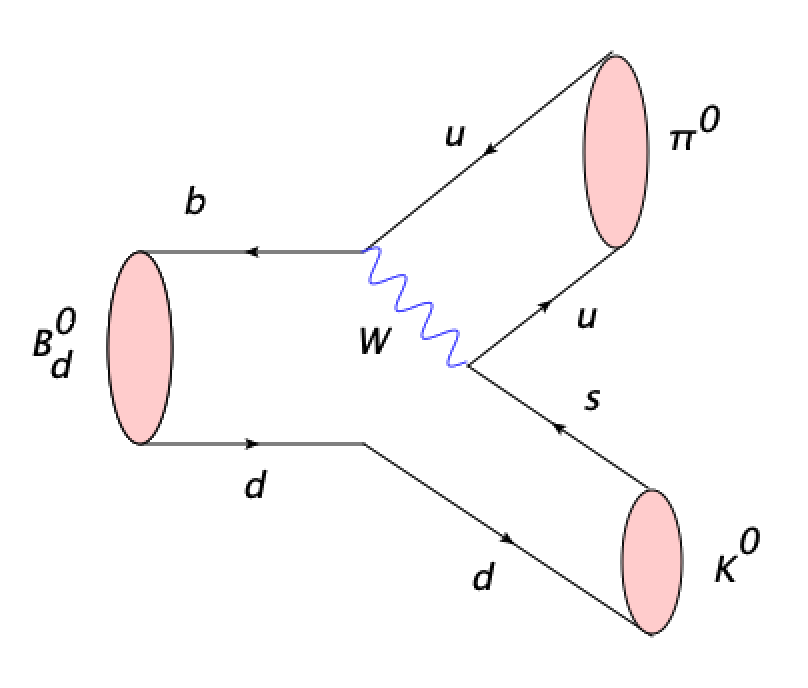}} \hspace{0.65cm}
	\subfloat[]{\label{fig:2}\includegraphics[width = 0.4\linewidth]{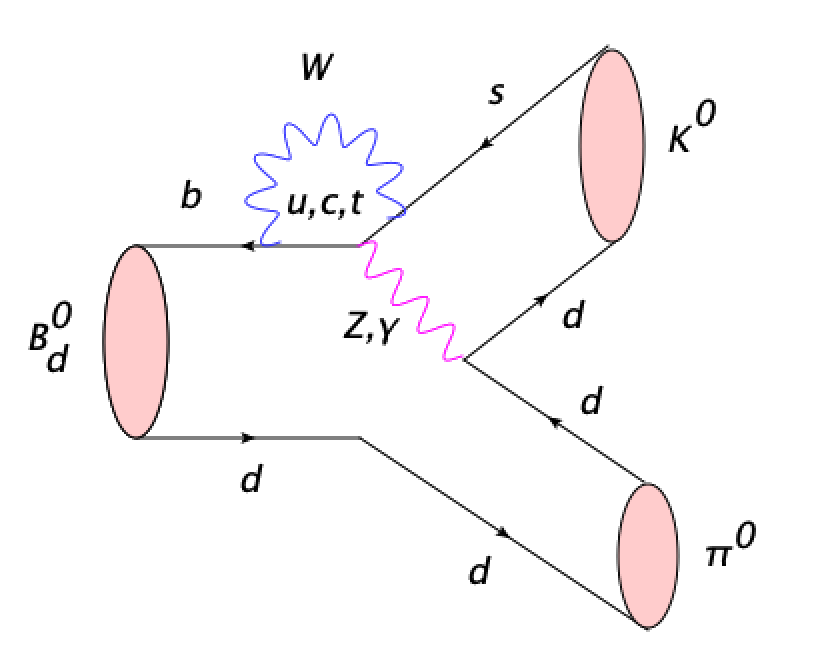}}\\
	\subfloat[]{\label{fig:1}\includegraphics[width = 0.35\linewidth]{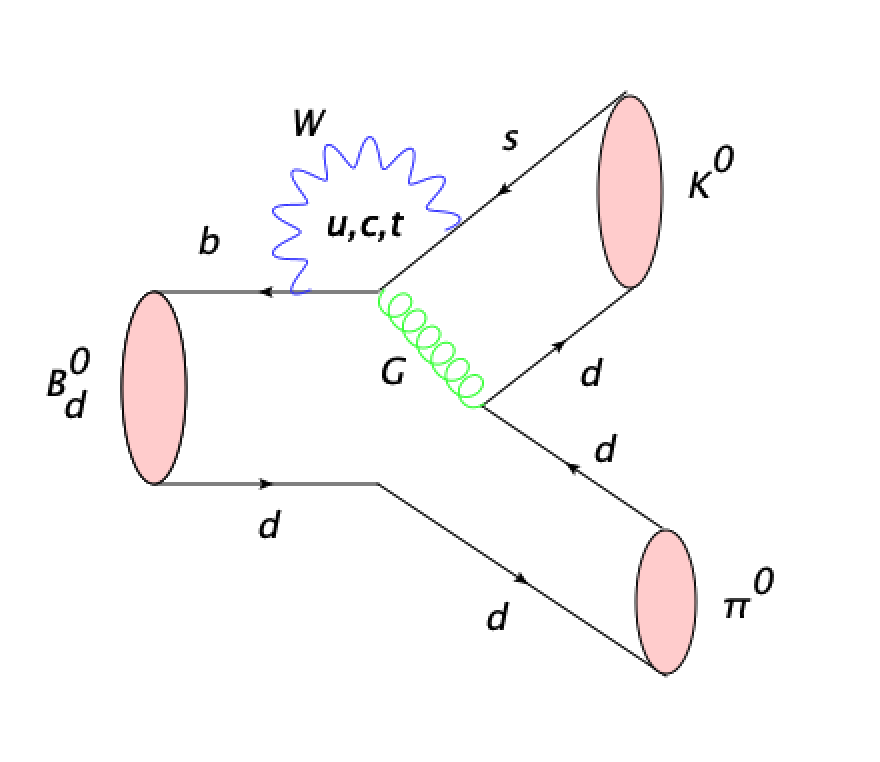}}\hspace{0.65cm}
	\subfloat[]{\label{fig:2}\includegraphics[width = 0.4\linewidth]{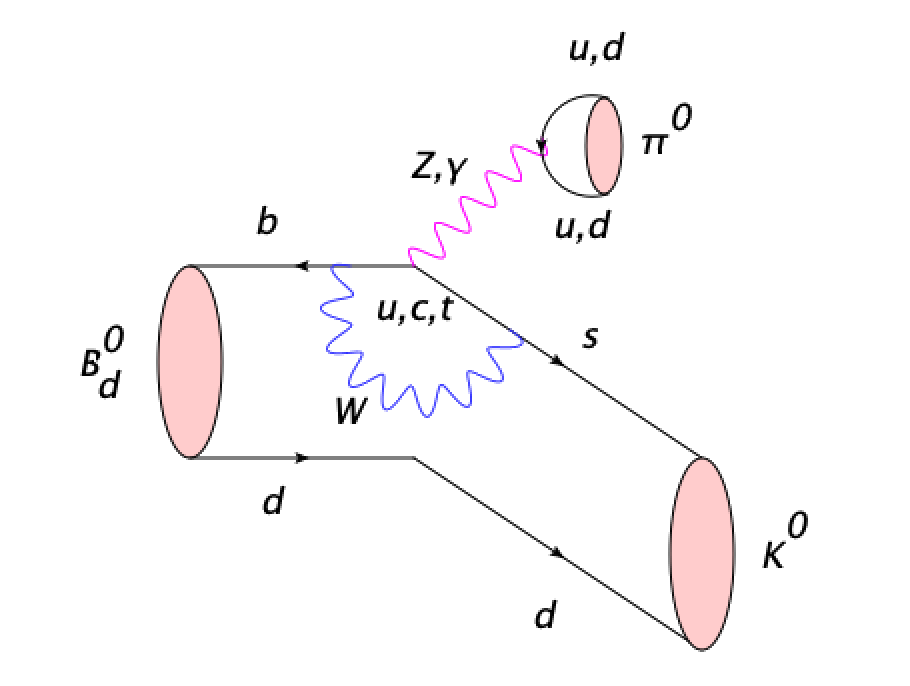}}
	\caption{The $B^0_d$ $\rightarrow$ $\pi^0 K_s$ Feynman diagrams: (a) colour suppressed tree, (b) colour suppressed EW penguin, (c) QCD penguin, (d) colour allowed EW penguin.}
	\label{fig:so}
\end{figure}

$B \rightarrow \pi K$ decays are non-leptonic decays, thus theoretically not clean, due to the presence of hadronic matrix elements. They are dominated by gluonic (QCD) loop diagrams (penguins). Contributions at tree level are suppressed, due to the tiny CKM matrix element $|V_{ub}|$. Electroweak (EW) penguins may also enter (at a comparable level to the tree amplitudes). A very promising decay channel is the $B^0_d$ $\rightarrow$ $\pi^0 K_s$. On Fig. \ref{fig:so}, as an example, we illustrate the topologies that contribute to this decay . 

The EW penguin topologies offer an avenue for new particles to enter  \cite{BFRS1}-\cite{London}. The New Physics (NP) contributions could involve new CP violation sources, which can be probed with the help of CP violating observables. The significance of the $B^0_d$ $\rightarrow$ $\pi^0 K_s$ decay is due to the fact that it is the only $B \rightarrow \pi K$ mode that has mixing-induced CP asymmetry, which arises from interference between $B^0 \-- \bar{B}^0$ mixing and their decay to the same final state. As we will see later, it plays a key role for testing the SM \cite{Fl99}. 

Here, we present an analysis of the correlation of the CP asymmetries of $B^0_d$ $\rightarrow$ $\pi^0 K_s$ and we use the current data to get SM with predictions. This is useful for the future analyses, especially for the Belle-II experiment, which has already started taking data. We check if there is any possibility for New Physics to be present. Moreover, we introduce two EW parameters, the $q$ and $\phi$, and we propose a new strategy in order to extract them. During our analysis, we use data and information from both the charged and the neutral $B \rightarrow \pi K$ decays as well as from the $B \rightarrow \pi \pi$ decays.

\section{Theoretical Framework}
The non-leptonic decays are challenging due to the presence of the hadronic matrix elements, which are very difficult to handle. The strong interactions are characterised by flavour symmetries, which link the amplitudes of $B \rightarrow \pi K$ with the  $B \rightarrow \pi \pi$, $B \rightarrow K K$ amplitudes. Thus, either there are hadronic amplitudes that get eliminated or being determined with the help of experimental data of the latter. In our studies, we keep the strong interactions theoretical assumptions as minimal as possible. Moreover, we use results of QCD factorization (QCDF) to include SU(3)-breaking corrections \cite{BeneNeu}.

First of all, we define the hadronic parameters:
\begin{align}
    r e^{i \delta} &= \left(\frac{\lambda^2 R_b}{1-\lambda^2}\right) \left[ \frac{T-(P_t - P_u)}{P_t-P_c} \right],\\
    r_c e^{i \delta_c} &= \left(\frac{\lambda^2 R_b}{1-\lambda^2}\right) \left[ \frac{T+C}{P_t-P_c} \right],\\
    \rho_c e^{i \theta_c} &= \left(\frac{\lambda^2 R_b}{1-\lambda^2}\right) \left[ \frac{P_t - P_u}{P_t-P_c} \right] \approx 0,\\
 \rho_n e^{i \theta_n} &= \left(\frac{\lambda^2 R_b}{1-\lambda^2}\right) \left[ \frac{C+(P_t-P_c)}{P_t-P_c} \right] = r_c e^{i \delta_c} - r e^{i \delta},
\end{align}
where $\lambda=|V_{us}|$ and $R_b=\left(1- \frac{\lambda^2}{2} \right) \frac{1}{\lambda} \left| \frac{V_{ub}}{V_{cb}} \right|$ (for numerical values of CKM elements, see \cite{PDG20}), $T$ and $C$, are normalized amplitudes that describe colour-allowed and colour-suppressed tree contributions and $P_u$, $P_c$, $P_t$ are penguin topologies. The terms  $ r e^{i \delta}$ and $ r_c e^{i \delta_c}$ are non-perturbative parameters, thus difficult to calculate. So, we calculate them using $B \rightarrow \pi \pi $ and SU(3) flavour symmetry \cite{BFRS1,BFL}:
\begin{align}
   r_c e^{i \delta_c} &= (0.17 \pm 0.06) e^{i(1.9 \pm 23.9)^{\text{o}}},\\
   r e^{i \delta} &= (0.09 \pm 0.03) e^{i(28.6 \pm 21.4)^{\text{o}}}.
\end{align}

As we know, regarding the $b \rightarrow s$ transitions, even though they are dominated by QCD penguins, the EW penguins also play an important role. So, describing these EW penguin effects, we are interested in two parameters: the parameter $q$, which provides a measure
of the strength of the EWPs with regard to tree topologies, and the CP violating weak phase $\phi$. Within SM, the phase $\phi$ vanishes and the q parameter can be determined using SU-(3) flavour symmetry of strong interactions.

We parametrise the EW penguin effects by:

\begin{equation}
\boldmath{ qe^{i \phi}e^{i \omega} =  \frac{\hat{P}_{EW}+\hat{P}^C_{EW}}{\hat{T}+\hat{C}}},
\end{equation}
where $\hat{T} $ denotes the color-allowed tree contributions, $\hat{C}$ the color-suppressed tree, $\hat{P}_{EW} $ the color-allowed EW penguin, $\hat{P}_{EW}^C$ the color-suppressed EWP and {$\omega$ the strong phase (which is quite small \cite{NRo} and vanishes in the SU-(3) limit), and we obtain the following relation \cite{Fl99,Bur98,NR,RSPJ}:
\begin{equation}
qe^{i \phi}e^{i \omega} = - \frac{-3}{2 \lambda^2 R_b} \left[ \frac{C_9(\mu)+C_{10}(\mu)}{C_1(\mu)+ C_2(\mu)} \right] R_q = (0.68 \pm 0.05) R_q,
\end{equation}
where $C_{i}(\mu)$ are perturbative Wilson coefficients \cite{Buch} and the $R_{q}$ is a parameter that describes SU-(3) breaking effects \cite{RSPJ}.
This ratio has been calculated for the SM values of $\phi = 0$. We are interested in checking whether there are any deviations from the SM values and in case there are, to see whether these deviations indicate physics beyond the SM.

As a next step, we consider the CP asymmetries in $B^0_d$ $\rightarrow$ $\pi^0 K_s$. The time-dependent $CP$ asymmetry is defined as:
\begin{align}
{A_{CP}(t) } &= { \frac{\Gamma (\bar{B}^0_d(t) \rightarrow \pi^0 K_s) - \Gamma (B^0_d(t)\rightarrow \pi^0 K_s)}{\Gamma (\bar{B}^0_d(t)\rightarrow \pi^0 K_s) + \Gamma (B^0_d(t)\rightarrow \pi^0 K_s)}} \nonumber \\
&= {A_{\pi^{0}K_s} \cos(\Delta M t) + S_{\pi^{0}K_s} \sin (\Delta M t)} ,
\end{align} 
where $\Delta M$ denotes the mass difference between heavy and light $B^0_d$ mass eigenstate. The term $A_{\pi^{0} K_s}$ describes the direct CP asymmetry and with the help of the amplitudes $|A(B^0_d \rightarrow \pi^0 K^0)| \equiv |A_{00}|$ and $|A(\bar{B}^0_d \rightarrow \pi \bar{K}^0)| \equiv |\bar{A}_{00}|$ is defined as:
\begin{equation}
{A_{\pi^{0} K_s} = \frac{|\bar{A}_{00}|^2 - |A_{00}|^2}{|\bar{A}_{00}|^2 + |A_{00}|^2}}.
\end{equation}
The term $S_{\pi^{0}K_s } $ plays the role of the mixing-induced $CP$ asymmetry (arising from interference between $B^0_d-\bar{B}^0_d$ mixing and decay processes of $B^0_d$,  $\bar{B}^0_d$ mesons into the same final state $\pi^0 K_s$) and with the help of the amplitudes is written as:
\begin{equation}
{S_{\pi^0K_s} = \frac{2 |A_{00} \bar{A}_{00}|}{|\bar{A}_{00}|^2 + |A_{00}|^2} \sin(2 \beta - 2 \phi_{\pi^0 K_s})},
\end{equation}
where we have for the angle: $2 \phi_{\pi^0 K_s}=\arg [\bar{A}_{00}A^{\ast}_{00}] \equiv \phi_{00}$ and the angle $2 \beta$ is the CP-violating $B^0_d-\bar{B^0_d}$ mixing phase $\phi_d$. The mixing-induced CP-asymmetry is an interesting probe of NP.

\section{State-of-the-art Analysis}
The direct and the mixing-induced CP asymmetry play a key role in our analysis. The two asymmetries are correlated as:
\begin{equation}
{S_{\pi^0K_s} = \sqrt{1 - A^2_{\pi^0 K_s}} \sin(\phi_d - \phi_{00})}.
\label{S-A}
\end{equation}
Since the direct asymmetry can be measured, we can make a prediction about the ${S_{\pi^0K_S}}$ asymmetry using Eq. \ref{S-A}. This requires the knowledge of the angle $\phi_{00}$. Therefore, an important question is how we determine this angle. 

\begin{figure}[t!]
	\centering
	\subfloat[]{\label{figa}\includegraphics[width = 0.45\linewidth]{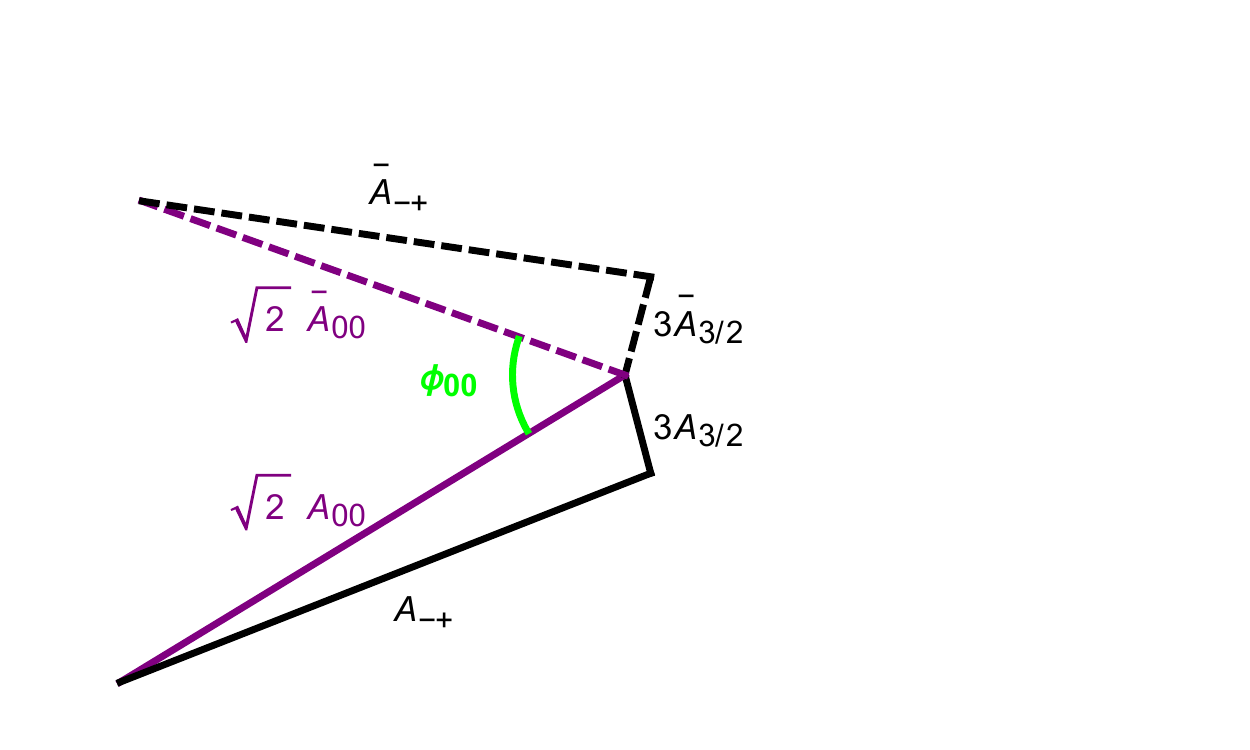}}
	\subfloat[]{\label{figb}\includegraphics[width = 0.5\linewidth]{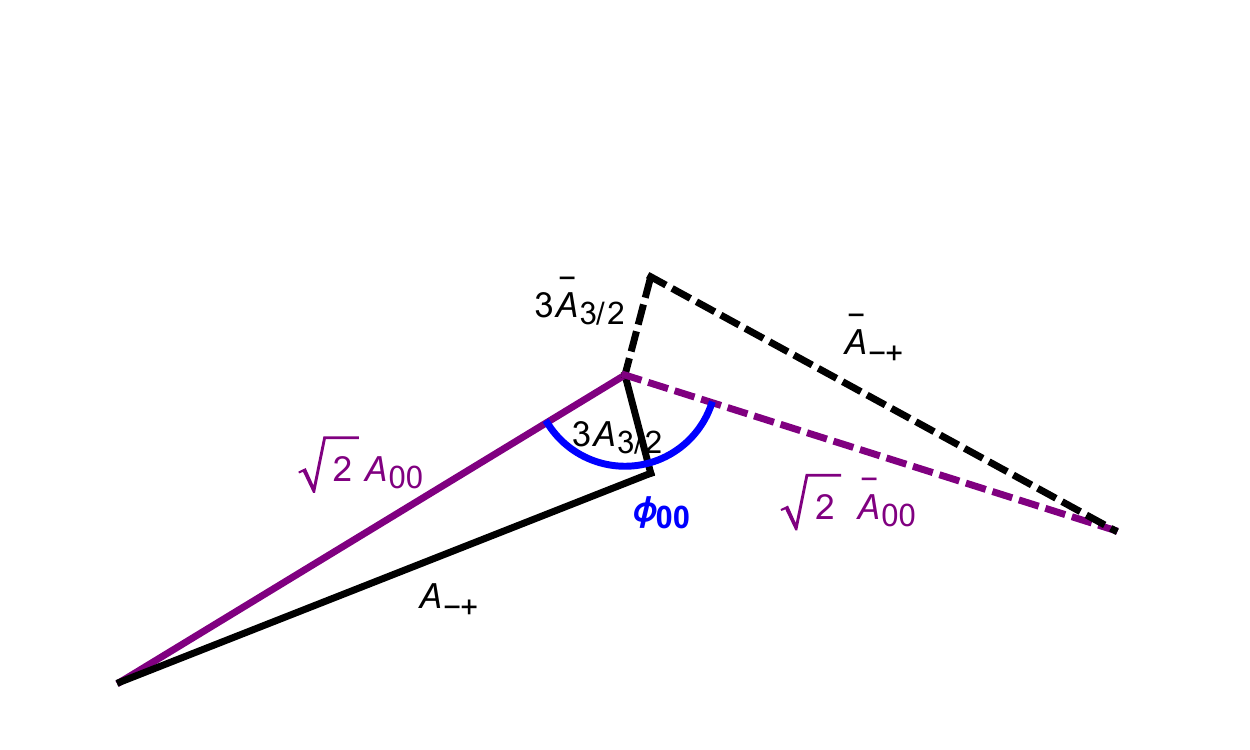}}\\
	\subfloat[]{\label{figc}\includegraphics[width = 0.45\linewidth]{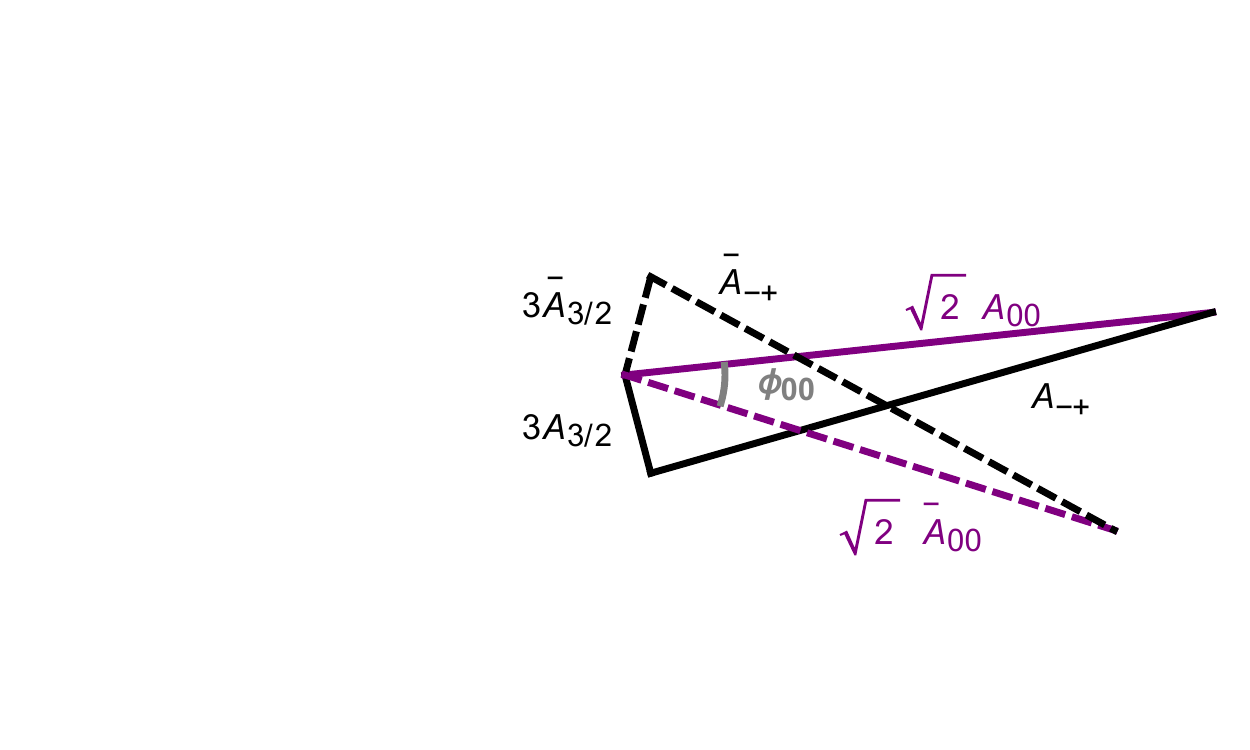}}
	\subfloat[]{\label{figd}\includegraphics[width = 0.5\linewidth]{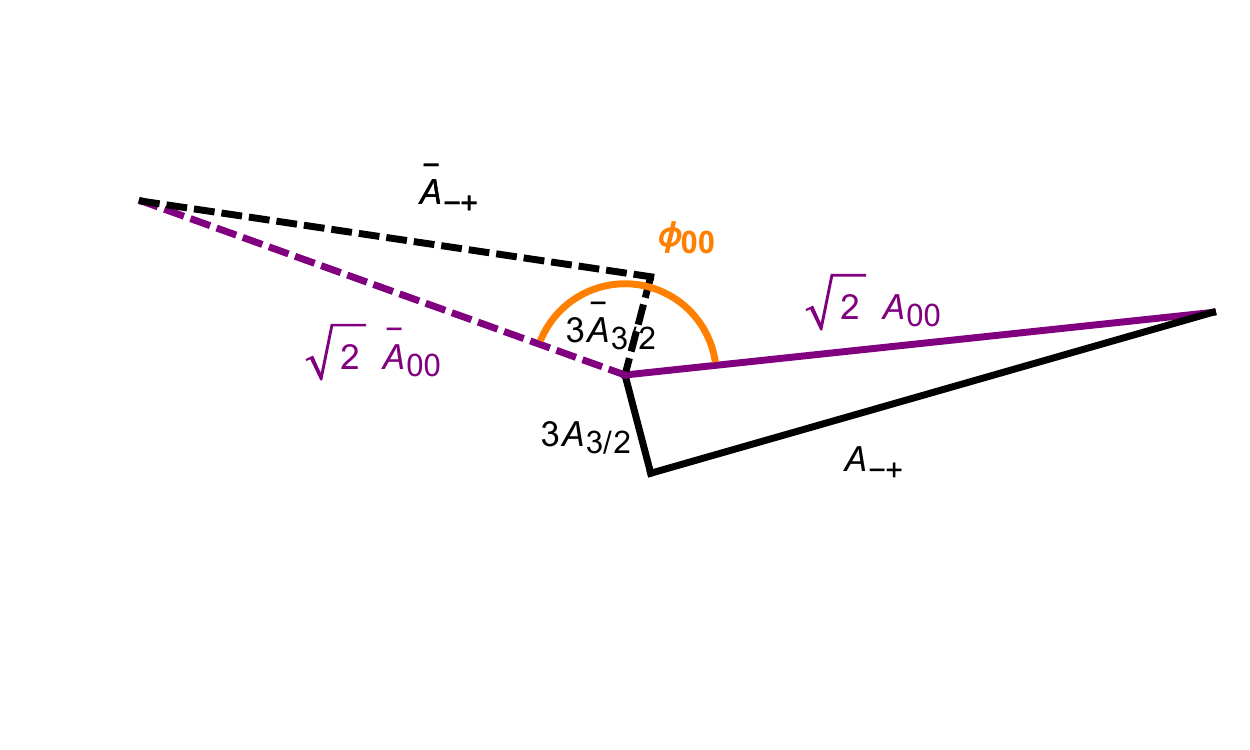}}\\
	\subfloat[]{\label{fig1a}\includegraphics[width = 0.48\linewidth]{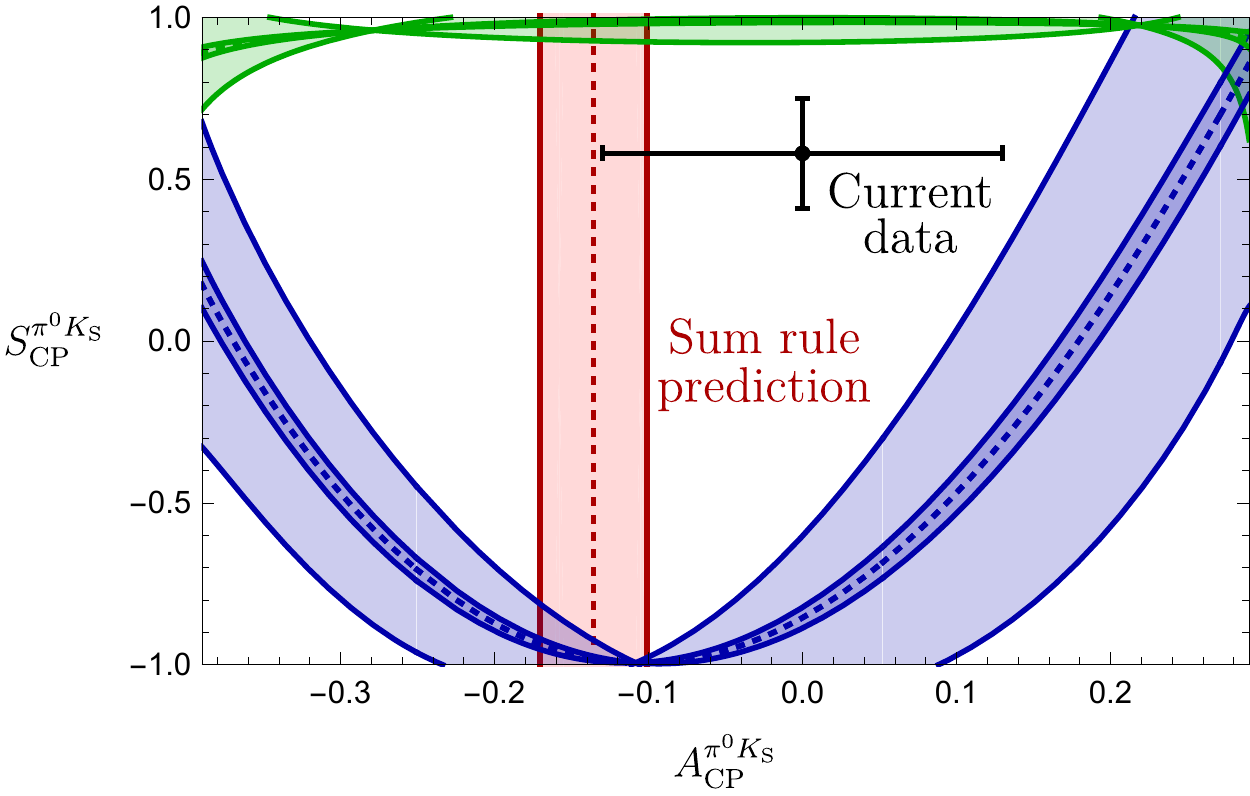}} \hspace{0.1cm}
	\subfloat[]{\label{fig1b}\includegraphics[width = 0.48\linewidth]{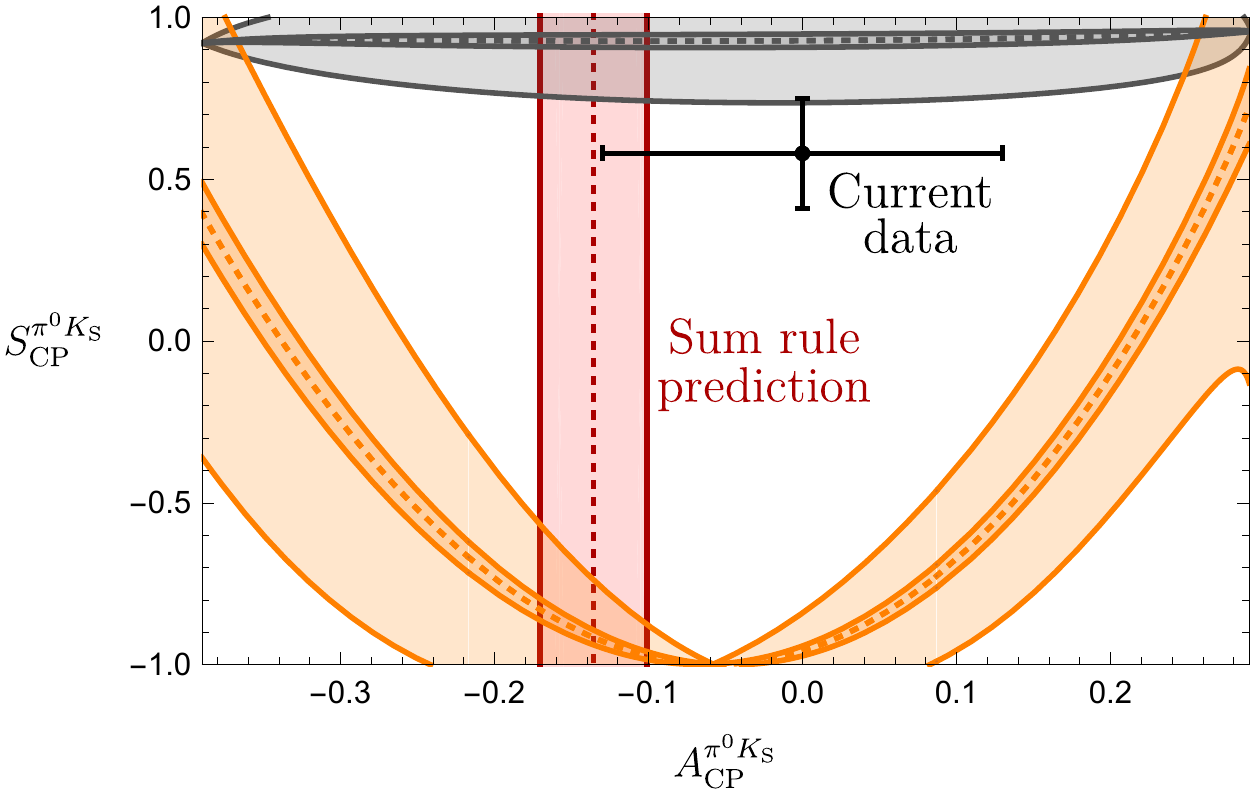}}\\
	\caption{(a)-(d): Amplitude triangles coming from the isospin relations, corresponding to the four different possible orientations. (e)-(f): Correlations between the direct and the mixing-induced CP asymmetries of $B^0_d \rightarrow \pi^0 K_s$. The values of the CP asymmetries for the current data are also indicated. The red vertical band is the sum rule prediction \cite{FJMV}.}
	\label{figtr}
\end{figure}

For this purpose, we use as a starting point the isospin relation \cite{NQ,GHLR}: 
\begin{equation}
\boldmath{3 A_{3/2}} = \boldmath{{A \left( B^{0}_d \rightarrow \pi^{-} K^{+} \right)} + {\sqrt{2} A \left( B^{0}_d \rightarrow \pi^{0} K^{0} \right)}} = -(\hat{T} + \hat{C}) (e^{i \gamma} - q e^{i \phi} e^{i \omega})
\label{iso}
\end{equation}
where $\gamma=\arg (V^{\ast}_{ub} V_{us})$. A similar relation can be written for the CP conjugate case, where the weak phases flip their sign while the strong phases keep the same sign. With the help of Eq. \ref{iso}, we are able to construct triangles in the complex plane. We can have $4$ different combinations between the amplitudes $A_{00}$ and $\bar{A}_{00}$, which means that we can get $4$ different angles $\phi_{00}$. As a result, we obtain four cases for the $S_{\pi^0K_s}$ asymmetry and this leads to four different branches in the $A_{\pi^0K_s} - S_{\pi^0K_s}$ plane \cite{FJMV}.

Fig. \ref{figtr}(a)-(d) illustrates the amplitude triangles that we obtain. The solid triangles refer to the $B^0_d \rightarrow \pi^0 K_s$ case while the dashed triangles to the CP conjugate. The triangles can flip around the axes $A_{3/2}$ and $\bar{A}_{3/2}$. Every angle $\phi_{00}$ indicates a different orientation and we denote each one of these angles with a different colour. Every $\phi_{00}$ corresponds to a contour in the $A_{\pi^0K_s} \-- S_{\pi^0K_s}$ plane (Fig. \ref{figtr}(e)-(f)) and the colour of the angle is the same with the colour of the contour that it refers to. Thus, the two bands in Fig. \ref{figtr}(e) illustrate the $S_{\pi^0K_s} - A_{\pi^0K_s}$ correlation that comes from $\phi_{00}$ of Fig. \ref{figtr}(a)-(b) while the bands in Fig. \ref{figtr}f arise from the angles in Fig. \ref{figtr}(c)-(d).  \cite{FJMV}. In Fig. \ref{figtr}(e)-(f), there are also the values of the CP asymmetries for the current data. The vertical red band on these plots refers to the prediction coming from the sum rule \cite{FJMV}.

Regarding the sum rule, making use of the CP-averaged branching ratios, we get the following relation \cite{Gro,GRos}:
\begin{eqnarray}
\label{eq:sum-rule-I}
\Delta_{\rm SR}^{({\rm I})} &=& A_\text{CP}^{\pi^\pm K^\mp} +  A_\text{CP}^{\pi^\pm K^0} \nonumber
\frac{\mathcal{B}(B^{+}\to\pi^{+}  K^0)}{\mathcal{B}(B^0_d\to\pi^-  K^+)} \frac{\tau_{B^0}}{\tau_{B^+}}\\ \nonumber
&\ & - A_\text{CP}^{\pi^0K^\pm} \frac{2 {\mathcal{B}(B^{+}\to\pi^0  K^+)}}{\mathcal{B}(B^0_d\to\pi^-  K^+)} 
\frac{\tau_{B^0}}{\tau_{B^+}} 
- A_\text{CP}^{\pi^0 K^0} \frac{2  \mathcal{B}(B^0_d\to\pi^0  K^0)}{\mathcal{B}(B^0_d\to\pi^-  K^+)} \\
&=& 0 + {\cal O}(r_{({\rm c})}^2,\rho_{\rm c}^2) \ .
\end{eqnarray}
This relation offers an interesting test of the SM. With the help of the hadronic parameters and the SM $q$ and $\phi$ values, the SM prediction gives  $\Delta_{\rm SR}^{({\rm I})}|_{{\text{SM}}}=-0.009 \pm 0.013$ while the value from the current data is $ \Delta_{\rm SR}^{({\rm I})}|_{{\text{SM}}}=-0.009 \pm 0.013$ \cite{FJMV}.

As we see, the amplitude triangles lead to a $4$-fold ambiguity. We are able to resolve this ambiguity (with the help of the strong phase ${\delta_c}$ \cite{RSPJ}) and to obtain only one solution, thus to pick the "correct" contour in the $A_{\pi^0K_s} - S_{\pi^0K_s}$ plane. The result is illustrated in Fig. \ref{sumpuzzle}(a) \cite{FJMV}. Therefore, the green band is the one that comes from the isospin analysis and again, the sum rule prediction and the current data result are shown. It is clear that there is tension with the SM prediction, which is what we call the $B \rightarrow \pi K$ puzzle.

\begin{figure}[b!]
	\centering
	\subfloat[]{\label{spuz}\includegraphics[width = 0.35\linewidth]{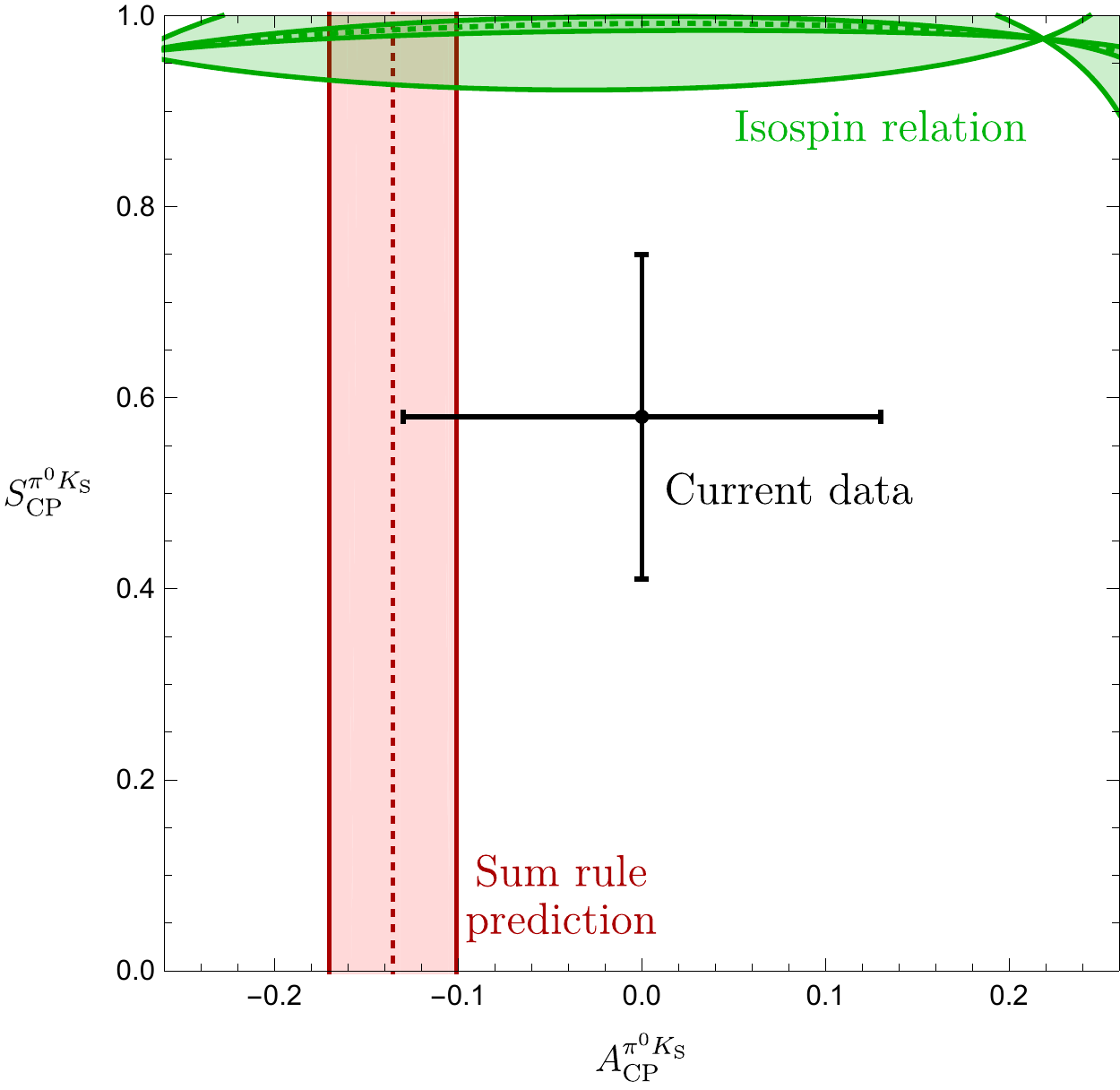}} \hspace{0.5cm}
	\subfloat[]{\label{puznew}\includegraphics[width = 0.35\linewidth]{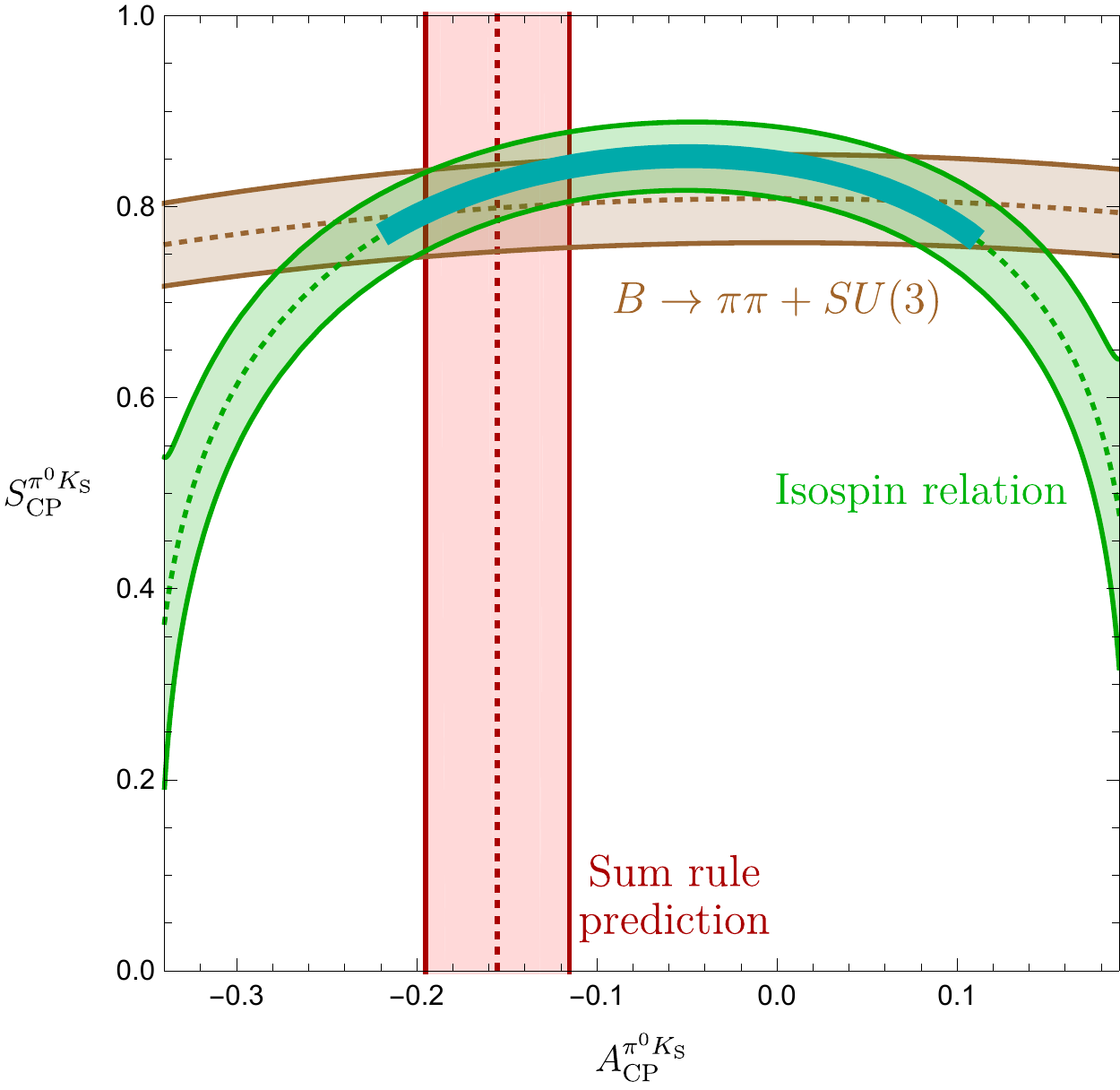}}
	\caption{(a) The $B \rightarrow \pi K$ puzzle. (b) Changing the data (lowering the $B^0_d \rightarrow \pi^0 K_s$ branching ratio), leads to a consistent picture with the SM \cite{FJMV}.}
	\label{sumpuzzle}
\end{figure}

\section{Can we resolve the {{$B \rightarrow \pi K$}} puzzle?}
An essential question is whether we can resolve the $B \rightarrow \pi K$ puzzle. Could it be resolved by a change of data or are there effects of NP? In the first case, it is very difficult to simultaneously fulfil all the constraints. It is interesting though to explore how the data should change, so that they agree with the SM. A good candidate would be the branching ratio, due to the large $B^0_d \rightarrow \pi^0 K^0$ experimental uncertainty. 

In our analysis, we found that lowering branching ratio's central value (2.5 $\sigma$) gives consistent picture with SM, as it is illustrated in Fig. \ref{sumpuzzle}(b) \cite{FJMV}. The green contour refers to the isospin relation while the brown contour gives the $S_{\pi^0 K_s} - A_{\pi^0K_s}$ correlation with the angle $\phi_{00}$ having been fixed with the help of the current data of $B \rightarrow \pi \pi$ and the SU-(3) symmetry.

However, this tension with the SM could be a hint of NP. As a very promising sector for NP signals is the EW penguin sector, this puzzle could be a signal for NP contributions in our decay. In this case, the parameters $q$ and $\phi$ would also be affected. An essential point is the sensitivity to new CP violation sources. NP scenarios with extra $Z^{'}$ boson could be introduced, which can provide links to anomalies in rare $B$ decays. 

\section{New Strategy to determine $q$ and $\phi$}
Having analysed the puzzling $B \rightarrow \pi K$ situation and with the picture that we get for the current data, which could indicate NP in the EW penguin sector, it is interesting to extract the $q$ and $\phi$ parameters from the data and to test the corresponding SM sector by comparing the results with the SM predictions. 

So far, we have worked using only the neutral $B \rightarrow \pi K$ system. Now, for the determination of the EW penguin parameters $q$ and $\phi$, we make use of the charged $B \rightarrow \pi K$ decays. In these charged decays, there is only direct CP violation and they follow an isospin analysis, which is similar to the relation that we presented for the neutral decays:
 \begin{equation}
    3 A_{3/2} = A(B^{+} \rightarrow \pi^{+} K^{0}) + \sqrt{2} A(B^{+} \rightarrow \pi^0 K^{+})
    =(\hat{T} + \hat{C}) (e^{i \gamma} - q e^{i \phi} e^{i \omega})
    \end{equation}
An important quantity for our calculations is the $R_c$ ratio of branching ratios of the following charged $B$ meson decays \cite{Bur98}:
   \begin{equation}
    R_c = 2 \left[ \frac{\mathcal{B}r(B^{+} \rightarrow \pi^0 K^{+})}{\mathcal{B}r(B^{+} \rightarrow \pi^{+} K^0)} \right] = 1.09 \pm 0.06
\end{equation}
where the numerical values \cite{FJMV} are the experimental values coming from the current data. 

We propose the following strategy in order to extract the parameters $q$ and $\phi$ \cite{FJMV}. Defining the angle $\Delta \phi _{3/2}$ as the difference between the $\phi_{3/2}$ (phase of the $A_{3/2}$ amplitude) and the $\bar{\phi}_{3/2}$ (phase of the $\bar{A}_{3/2}$ amplitude) and converting $A_{3/2}$ to the quantity $N$ as: $\sqrt{N} = 3 \left|{A_{3/2}}\right| / \left|{\hat{T}+\hat{C}} \right|$, we obtain the following relations for $q$ and $\phi$:
\begin{equation}
{q = \pm \sqrt{N + 1 -2c \cos \gamma -2s \sin \gamma}}
\end{equation}
\begin{equation}
{\tan \phi = \frac{\sin \gamma - s}{\cos \gamma - c }}
\end{equation}
where the quantities $c$ and $s$ are defined as: 
\begin{equation}
c = \pm \sqrt{N} \cos \left( \frac{\Delta \phi_{3/2}}{2} \right) {\text{\ \ \ \ and\ \ \ \ }} s = \pm \sqrt{N} \sin \left( \frac{\Delta\phi_{3/2}}{2} \right).
\end{equation} 
Using these relations, we are able to derive contours in the $q-\phi$ plane. 

Fig. \ref{qphi} shows the picture that we get for the current data \cite{FJMV}. The four contours arise from the isospin analysis (keeping the same colour coding that we used before). The cyan dotted line complements the analysis with the quantity $R_c$. Making use of the hadronic parameters, $R_c$ can be written in the form:
\begin{equation}
 R_c = 1- 2 r_c \cos \delta_c (\cos \gamma - q \cos \phi) + \mathcal{O}(r_c^2),
\end{equation}
which allows us to convert $R_c$ into another contour. The isospin analysis contours are in a very good agreement with the $R_c$ contour. The SM values of $q$ and $\phi$ with the corresponding errors are also indicated in the plot.

\begin{figure}[t!]
	\centering
	\includegraphics[width = 0.4\linewidth]{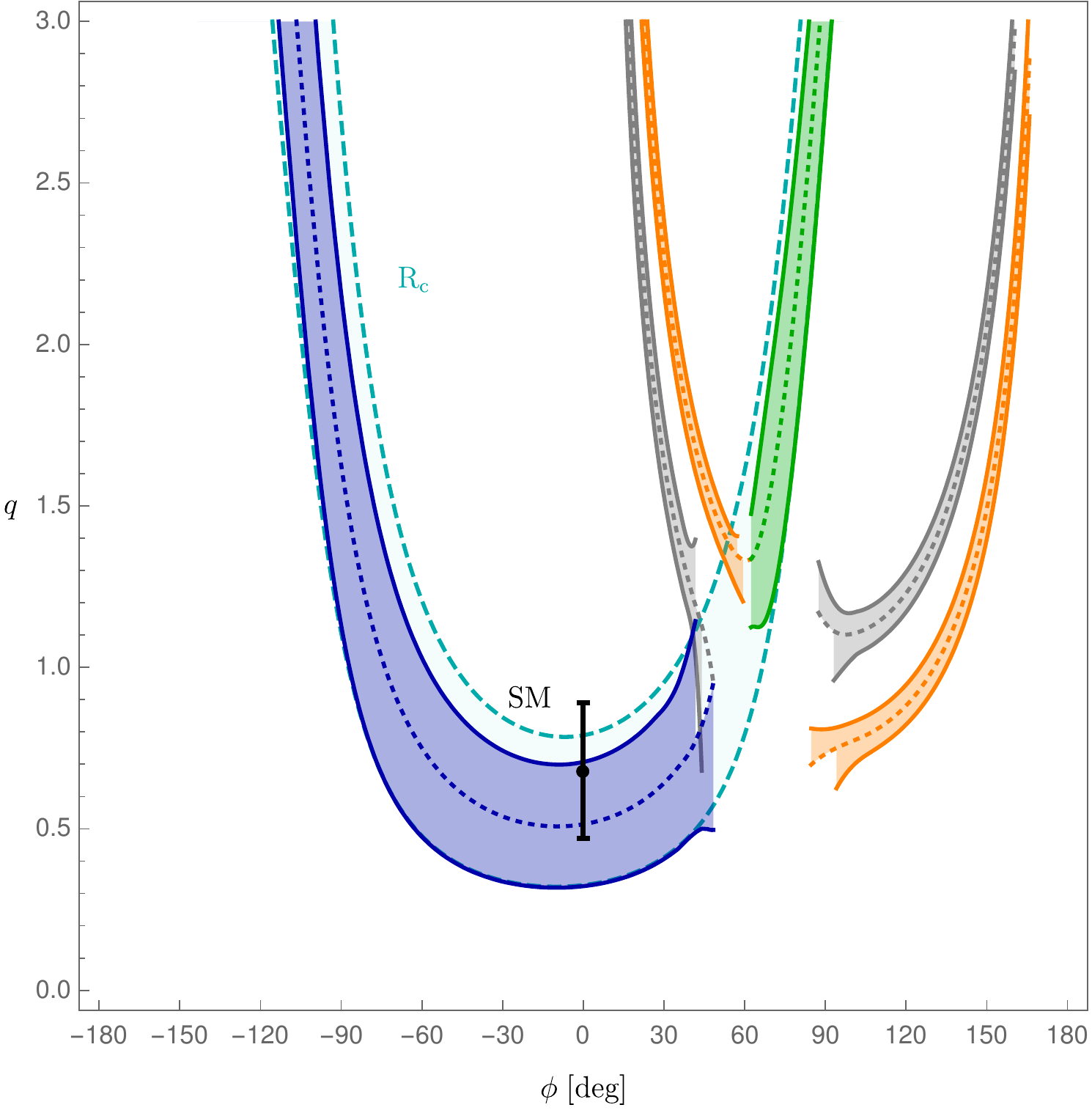} 
		\caption{Contours in the $q-\phi$ plane including the error bands. Four contours arise from isospin relation and the cyan dotted one is derived with the help of $R_c$. The SM value is also shown \cite{FJMV}.}
	\label{qphi}
\end{figure}

\section{Constraints on $q$ and $\phi$ and future scenarios}
We are interested in getting a sharper picture regarding the $q$, $\phi$ determination. For this purpose, we work again with the neutral $B \rightarrow \pi K $ system and we utilise the mixing-induced CP violation $S^{\pi^0 K_s}_{CP}$. Using $S^{\pi^0 K_s}_{CP}$, we are able to extract the phase $\phi_{00}$. Using the current measurements, the value of $\phi_{00}$ is calculated:
\begin{equation}
\phi_{00}=(7.7 \pm 12.1)^{\text{o}} \  \cite{FJMV}.
\end{equation} 

\begin{figure}[h!]
	\centering
	\subfloat[]{\label{spu}\includegraphics[width = 0.37\linewidth]{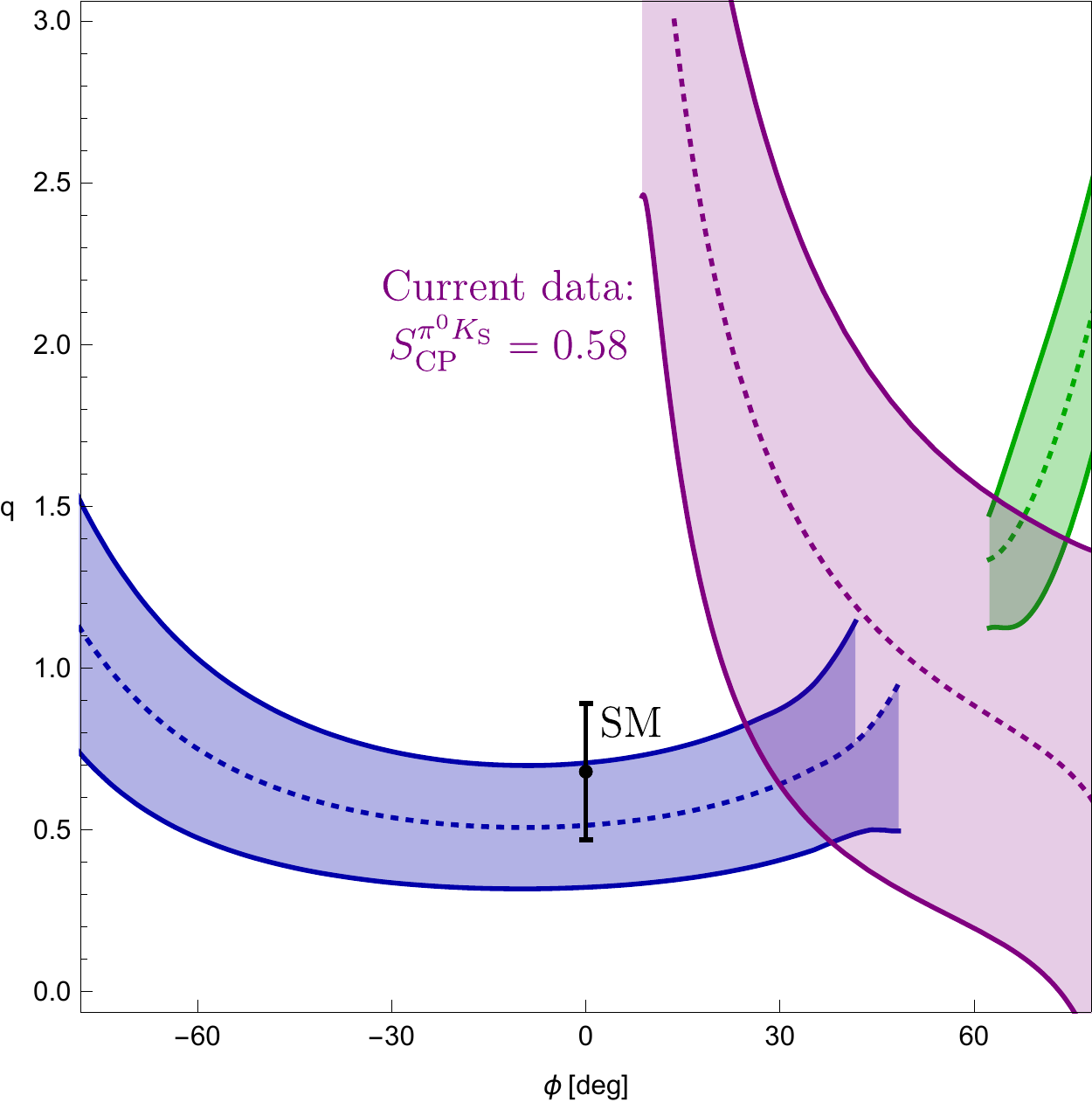}} \hspace{0.6cm}
	\subfloat[]{\label{puz}\includegraphics[width = 0.37\linewidth]{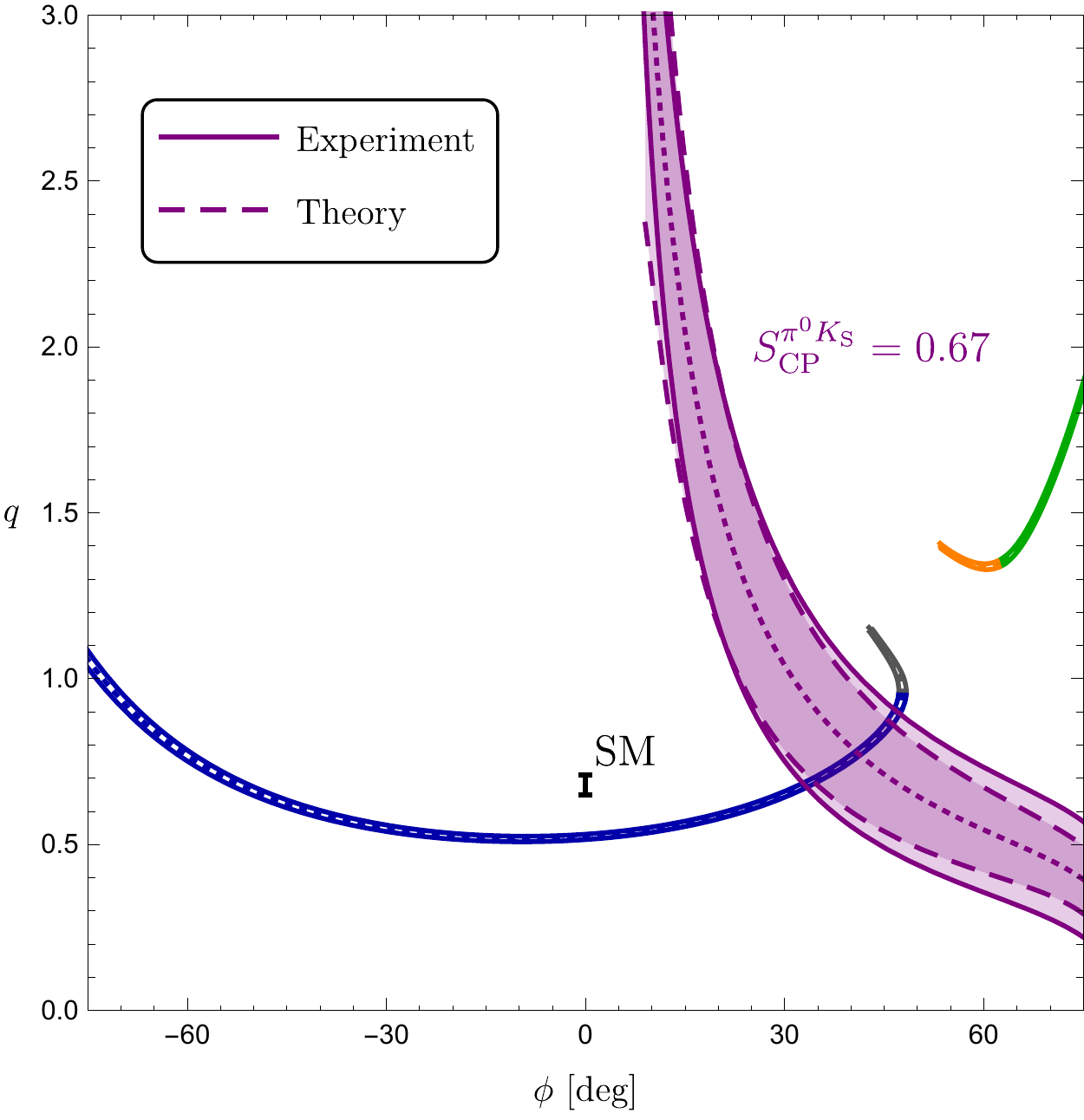}}
	\caption{(a) Constraints on $q$, $\phi$, utilising information from the mixing-induced CP asymmetry $S^{\pi^0 K_s}_{CP}$. (b) Future scenario. The experimental and theoretical errors are given separately \cite{FJMV}.}
	\label{future}
\end{figure}

With the help of the hadronic parameters, mixing-induced CP violation can be converted into contour on the $q- \phi$ plane. This contour is shown as purple in Fig. \ref{future}a. Contributions from color-suppressed EW penguins can also be included using data. The blue and the green contour on the plot are again the isospin analysis contours. The SM point is also given.

Our next step is to apply our strategy to a future scenario and illustrate it. We consider a scenario for ${S^{\pi^0 K_s}_{CP}}$ measurement. We assume an uncertainty, same as the uncertainty of the direct asymmetry, which can be reached at the end of Belle II \cite{Abe}. The constraints coming from ${S^{\pi^0 K_s}_{CP}}$  asymmetry and the isospin relations are shown on Fig. \ref{future}b (following the same colour coding as before). Errors coming from the experiment are denoted by the solid lines while the error bands that correspond to theory predictions are indicated by the dashed line. The important result is that the experimental precision and theory can be matched.

\section{Conclusion}
The $B \rightarrow \pi K$ puzzle remains an intriguing problem over the years, that can either confirm the SM or reveal NP. Here, using the current experimental data, we performed a state-of-the-art analysis providing an updated picture of the current situation. 

The correlation between the mixing induced and the direct CP asymmetries revealed a tension in the data, that got stronger in comparison with previous results, which shows something very interesting: either the data should "move" to become consistent with the SM or NP might be present. EW penguins, with the characteristic parameters $q$ and $\phi$, provide an avenue, which allows new particles to enter, thus it is an interesting section to search for NP.

We proposed a new strategy to determine $q$ and $\phi$, using both neutral and charged $B \rightarrow \pi K$ decays. Deriving expressions for these parameters, we calculated contours in the $\phi - q$ plane. We obtained a more complete picture utilising information from the ${S^{\pi^0 K_s}_{CP}}$  asymmetry.  We applied this strategy in a future scenario and we showed that the experimental errors can match our theoretical ones. 

The implementation of our new method at Belle II will clarify the things either by confirming again the SM model or by announcing NP
with new CP violation sources. Thus, data from Belle II and LHCb upgrade will allow exciting new opportunities. In case of NP establishment, new particles like $Z'$ bosons could be connected to anomalies in rare decays $B \rightarrow K^{\ast} \ell^{+} \ell^{-}$, leading to fascinating new discoveries.

\section*{Acknowledgements} 
I would like to thank R. Fleischer, R. Jaarsma and K.K. Vos for the enjoyable collaboration during the preparation of the paper \cite{FJMV}, which is the main paper that this talk is based on. I would also like to thank the organisers of the "Workshop on Connecting Insights in Fundamental Physics: Standard Model and Beyond - Corfu2019" for giving me the opportunity to give this talk.

\end{document}